\documentclass[aps,prb,twocolumn,superscriptaddress,floatfix,citesort,showpacs]{revtex4}
\usepackage{epsfig}
\begin{document}

\title{Mode Coupling and Dynamical Heterogeneity in Colloidal Gelation:\\ A Simulation Study}
\author{Antonio M. Puertas}
\affiliation{Group of Complex Fluids Physics, Department of Applied Physics, University of Almeria, 04120 Almeria, Spain}
\author{Matthias Fuchs}
\affiliation{Fachbereich Physik, University of Konstanz, D-78457 Konstanz, Germany}
\author{Michael E. Cates}
\affiliation{School of Physics, The University of
Edinburgh, Kings Buildings, Mayfield Road, Edinburgh EH9 3JZ, UK}
\date{\today}

\begin{abstract}
We present simulation results addressing the dynamics of a colloidal system with attractive interactions close to gelation. Our interaction also has a soft, long range repulsive barrier which suppresses liquid-gas type phase separation at long wavelengths. The new results presented here lend further weight to an intriguing picture emerging from our previous simulation work on the same system. Whereas mode coupling theory (MCT) offers quantitatively good results for the decay of correlators, closer inspection of the dynamics reveals a bimodal population of fast and slow particles with a very long exchange timescale. This population split represents a particular form of dynamic heterogeneity (DH). Although DH is usually associated with activated hopping and/or facilitated dynamics in glasses, the form of DH observed here may be more collective in character and associated with static (i.e., structural) heterogeneity. 
\end{abstract}

\maketitle

\newpage

\section{Introduction}
Colloidal systems have an important experimental role as model materials for studies of the glass transition \cite{pusey,introd}. In particular, it is possible to approach the limit of hard sphere particles, and then deviate from this in a controlled manner by the addition of interactions, such as the depletion interaction mediated by added polymer \cite{asakura54}. Recently this has allowed the role of short range attractions to be probed experimentally \cite{pham02,eckert02}; that work has confirmed a scenario of re-entrant melting (on addition of a weak short ranged attraction) followed by re-freezing into an attraction-driven glass. This scenario was first suggested by MCT (mode coupling theory) calculations \cite{bergenholtz99,fabbian99,dawson01}. Its confirmation in experiments represents strong evidence that MCT, while admittedly deficient in its neglect of activated local dynamics \cite{gotze91,gotze92,gotze99}, does capture some of the important collective dynamics of the colloidal arrest transition. 

Within MCT, these collective dynamics are dominated by colloidal interaction forces (repulsive caging or attractive bonding) rather than details of the local diffusivity. For this reason, when simulating such systems, many-body hydrodynamic coupling is frequently ignored, replacing the full hydrodynamic interactions with local drag (Brownian dynamics: BD). Moreover, if the local dynamics is truly unimportant, one may even replace Brownian dynamics with Newtonian (i.e., molecular dynamics: MD) which further saves computer time. In effect the solvent has then been replaced by a vacuum. This brings the simulated system very close to that of atomic or molecular glasses, although the interaction parameters are typically somewhat different: a shorter range of attraction is common in colloids.

Here we report various MD simulation results on model colloidal systems with short-range attractions, close to the arrest transition (gelation). These extend the work presented by ourselves in a series of earlier papers \cite{puertas02,puertas03,puertas04} which revealed a somewhat paradoxical picture of dynamics in these systems. We found \cite{puertas02,puertas03} that MCT is quantitatively successful at predicting correlation functions and other ensemble-averaged properties close to arrest; this finding offers strong evidence for the physical assumptions of collective rather than activated dynamics underlying MCT. Yet, when we looked at individual particle motions \cite{puertas04}, the behavior is far more complicated than we expected, revealing a strongly bimodal distribution of particle displacements. These can be resolved, over prolonged time scales, into two separate populations of fast and slow particles (with a very sluggish exchange between these). Moreover the fast and slow particles are highly correlated in space; at any instant, the dynamics resembles a near-frozen gel of slow particles with fast particles moving around and through channels in the gel \cite{puertas04}. Part of this picture may result from a long-range repulsive contribution to the chosen interaction potential whose role is explained below; nonetheless, the results can shed important light on the nature of collective and local motions in arresting systems.

The behavior that we have observed represents a specific form of dynamical heterogeneity (DH), related to, but distinct from, the DH found for glasses arrested by caging \cite{kob97,donati99a,donati99b,weeks00,weeks02}. The latter observations of DH tend to support a `facilitated dynamics' picture of the glass transition in which regions of high mobility interact nonlinearly to give a super-activated temperature dependence of the overall mobility. Models of this type have been studied for several years \cite{kob95} and lately raised to a higher level of predictive power by Chandler and coworkers \cite{chandler}. However, these approaches generally model a dynamics in which a frozen matrix (often represented by a lattice) is relaxed progressively as the heterogeneities visit different neighborhoods. Such models do not yet capture the rather rich picture observed in our simulations; for example, fast particles are found preferentially at the surface of the frozen component and individual bonds between fast particles have shorter lifetimes than bonds between slow ones \cite{puertas04}. These features probably cannot be understood within a lattice DH model but would require explicit treatment of off-lattice dynamics in the presence of short range bonding.

An important question is why the presence of such heterogeneities does not destroy the agreement with MCT found in the ensemble-averaged dynamics, as measured in correlation functions \cite{puertas03}. We do not yet have a clear answer to this question. However, we note that the heterogeneous dynamics observed, once the fast and slow components are identified, is correlated with structural heterogeneity; and this is (at least partly) collective in origin. The heterogeneity is therefore not simply attributable to the activated `local hopping' channel which was long ago recognized to be missing from MCT \cite{gotze91,gotze92}. Indeed, some aspects of this population-level dynamic heterogeneity may have little to do with the physics of activation (although that does seem to control a slow exchange between populations that we observe; see Section \ref{DH} below).

If the observed dynamic heterogeneity reflects in part collective rather than activated processes, then its presence in a system whose overall properties (as measured by decay of globally averaged correlators) are well described by MCT is more easily explicable. Such dynamics could arise, for example, by an interplay between MCT-like arrest and incipient liquid-gas like phase separation. Macroscopic phase separation is prevented, in our simulations, by a barrier in the potential \cite{puertas02}. In real life, even for systems quenched beyond the relevant spinodal, it could be prevented instead by the extremely hindered separation dynamics that arise if one of the two coexisting phases is nonergodic. Incipient fluctuations of the same physical character should also exist close to the spinodal but within the stable phase. (In our own work we do not see microphase separation but do not rule out an incipient or frustrated form of this.)

The interplay of phase separation and arrest has recently been discussed for colloidal systems at relatively low density, where some sort of iterative renormalisation of the MCT parameters is required \cite{kroy03}. At higher colloid density, as we study here, MCT could in principle capture some of this interplay unaided; after all, the input for MCT is the static structure factor $S(q)$ which contains, as $q\to 0$, information about proximity to the liquid-gas spinodal. (In our system where a barrier in the potential suppresses this spinodal, $S(q)$ contains instead information about any incipient microphase separation through the `pre-peak' at low $q$.) However, we emphasise that in other aspects of the observed dynamics, particularly connected with interchange between fast and slow populations, MCT could well offer only limited insight, as discussed below. A somewhat complicated picture involving activation and/or facilitated dynamics, perhaps coupled to structural inhomogeneity and collective dynamical modes, may ultimately be required to model the system studied here. 

In what follows, we first recall in Section \ref{details} the details of our simulations. We then present new data that offers further quantitative support for the picture of dynamics close to colloidal gelation that was outlined above. In Section \ref{MCT} we add to our MCT-inspired analysis \cite{puertas03} by presenting new results for the coherent autocorrelator, its wavevector-dependent nonergodicity parameter $f_q$, and its wavevector-dependent terminal time $\tau_q$. In Section \ref{DH} we add to our population-based analysis of DH\cite{puertas04} with additional quantitative measures of the differing local environments of fast and slow particles, based on the number of neighbors of each type and on measures of percolation. In Section \ref{conc} we give our conclusions.

\section{Simulation details}
\label{details}
We have performed computer simulations of a system composed of soft core polydisperse particles, with a short-range
attraction given by the Asakura-Oosawa potential \cite{asakura54}, modeling a mixture of
colloids with non-adsorbing polymers. Phase separations have been avoided, in order to have full access to the whole composition space. Crystallization is inhibited by polydispersity, and liquid-gas demixing by a long range repulsive barrier. The gel transition is generally approached from the fluid side, by increasing the polymer fraction within the fluid phase. 

The total interaction has three parts, a core repulsion, a short range attraction and a repulsive barrier. The core repulsion is given by:

\begin{equation}
V_{sc}(r)\:=\:k_BT \left(\frac{r}{a_{12}}\right)^{-36}
\end{equation}

\noindent where $a_{12}=(a_1+a_2)$, with $a_1$ and $a_2$ the radii of the
interacting particles. Particle sizes are distributed according to a flat
distribution of half-width $\delta=0.1 a$, where $a$ is the mean radius;
$a_i \in [0.9a,1.1a]$. The attraction induced by the polymers, extended to
take polydispersity into account, reads \cite{asakura54,dijkstra99,mendez00}:

\[ V_{AO}(r) \:=\: -k_BT \phi_p \left\{\left[\left(\bar{\eta}+1\right)^3 -\frac{3r}{4\xi} \left(\bar{\eta}+1\right)^2+\frac{r^3}{16\xi^3}\right]+
\right.\]
\begin{equation}\label{pot} 
\left.+\frac{3\xi}{4r} \left(\eta_1-\eta_2\right)^2 \left[\left(\bar{\eta}+1\right) -\frac{r}{2\xi} \right]^2\right\}
\end{equation}

\noindent for $2(a_{12}+\xi/5) \leq r \leq 2(a_{12}+\xi)$ and $V_{A0} = 0$ for larger
distances. Here, $\eta_i=a_i/\xi$; $\bar{\eta}=(\eta_1+\eta_2)/2$, and
$\phi_p$ is the volume fraction of the polymer. The range of
the potential is given by $\xi$, the polymer size, and its strength is
proportional to $\phi_p$. At short distances, for computational efficiency this potential is replaced
by a steep parabola with the minimum at $r=2 a_{12}$; this ensures that the total
interaction potential retains a quadratic minimum very close to $r=2 a_{12}$.

The long range repulsive barrier inhibits liquid-gas separation by raising the energy of the dense phase. It has the form:

\begin{equation}
V_{bar}(r)\:=\:k_BT\left\{\left(\frac{r-r_1}{r_0-r_1}\right)^4-2\left(
\frac{r-r_1}{r_0-r_1}\right)^2+1\right\}
\end{equation}

\noindent for $r_0\leq r \leq r_1$ and $V_{bar}(r) = 0$ otherwise. The limits of the
barrier were set to $r_0=2(a_{12}+\xi)$, and $r_1=4a$, and its height is
only $1 k_BT$. This energy is equal to the attraction strength at
$\phi_p=0.0625$. As described in the introduction this leads to a low-$q$ peak in the structure factor but this is relatively mild (always lower than the main peak) and various tests for microphase separation gave negative results \cite{puertas03}. However, as previously discussed we do not rule out some effect of frustrated or incipient microphase separation on the dynamics.

The resulting total interaction potential, $V_{tot}=V_{sc}+V_{AO}+V_{bar}$, is analytic for all distances and can be used with simple molecular dynamics algorithms. Polydispersity in size causes differences in the interaction between different pairs of particles; apart from the obvious differences in the core radii the attractive well depth also increases with particle size, the maximal variation being \cite{puertas03,puertas04} $1.25\,k_BT$.

The simulated system is composed of $1000$ particles, length is measured in units of the average radius, $a$, and time in units of $\sqrt{4a^2/3v^2}$, where the
thermal velocity, $v$, was set to $\sqrt{4/3}$. Equations of motion were
integrated using the velocity-Verlet algorithm, in the canonical ensemble
(constant NTV), to mimic the colloidal dynamics, with a time step equal to
$0.0025$. Every $n_t$ time steps, the velocity of the particles was
re-scaled to assure constant temperature. No effect of $n_t$ was observed
for well equilibrated samples. Equilibration of the samples was tested by
the trends of the energy of the system and other order parameters
\cite{puertas03}, and by the time-translation invariance of the correlation functions. The range of the attraction is set to $2\xi=0.2a$, and the
control parameters are (i) the density of colloids, reported as volume fraction,
$\phi_c=\frac{4}{3}\pi a^3 \left(1+\left(\frac{\delta}{a}\right)^2\right)
n_c$, with $n_c$ the colloid number density, and (ii) the polymer volume fraction,
$\phi_p$, which measures the attraction strength.

Because of the dynamical heterogeneities observed, care is needed in gaining adequate statistics over quasi-independent realisations. For every state point, different systems were equilibrated by
quenching from hard spheres to the desired $\phi_p$. For the slowest evolving states ($\phi_p=0.405$, $0.41$, $0.415$ and $0.42$) ten different quenches were
performed and $100$ different time origins were considered for computing correlation functions, all of which lay beyond the initial equilibration period. The reported correlation
functions for these states are thus, in effect, averaged over $1000$ independent
realizations. The equilibration period was as long as $5\times 10^4$ time units in the slowest cases; this was enough to give time-translation invariant correlators but nonetheless may not be long compared to the timescale of exchange of particles between  fast and slow populations \cite{puertas04}.
The issue of whether samples are ``fully'' equilibrated is therefore partly a matter of interpretation -- as is often the case in systems approaching an arrest transition, and always the case within the gel phase itself.

\section{Results and Discussion}

\begin{figure}
\psfig{file=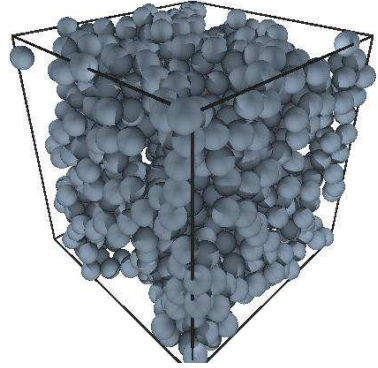,width=7.5cm}
\caption {\label{snapshot} Snapshot of the system close to the gel transition: $\phi_c=0.40$ and $\phi_p=0.42$.}
\end{figure}

\begin{figure}
\psfig{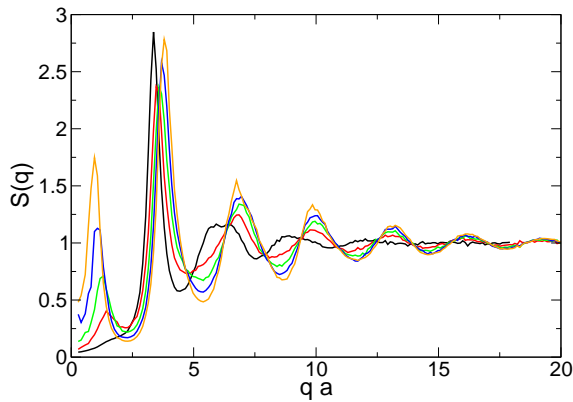}
\caption {\label{sq} Structure factor for different states on the isochore $\phi_c=0.40$. Approaching the gel transition: $\phi_p=0$, $0.20$, $0.35$ and $0.42$.}
\end{figure}

The system under study has been specially devised to allow us full access to the whole $(\phi_c,\phi_p)$ plane, eliminating both crystallization and liquid-gas demixing. However, the interplay between the long range repulsive barrier and the strong short range attractions induces a somewhat heterogeneous structure in the system, with voids and tunnels (see Fig. \ref{snapshot}), which minimizes the potential energy. Such a heterogeneous system is characterized by a peak at low (but finite) wavevector in the structure factor, $S(q)$. In our system, such a pre-peak forms as the attraction strength is increased, as observed in Fig. \ref{sq}, showing the build up of the heterogeneous structure. 

A low angle peak has been also noticed in experiments on colloidal gelation, moving to lower wavevectors as time proceeds, until it finally arrests \cite{segre01,haw95}. Despite the apparent similarity between both peaks, their origin is quite different; whereas the experimental one is due to the arrest of liquid gas demixing by gelation, the peak in the simulations is an equilibrium feature, caused by the long range repulsion. Nonetheless, the effect of these
two mechanisms on dynamics within the final structure could be similar, and we believe that MD studies of the system with barrier offer the closest numerical analogue currently available for such experiments. (If no barrier were used, because the simulation is small compared to a real system, substantial phase separation on the length scale of the simulation box would occur
quite rapidly \cite{puertas03,cates04}.)

Under certain circumstances, long range repulsive barriers are known to cause microphase separation \cite{sear99,sear99b,canongia02}. In that case, the regions of high and low particle density arrange in a periodically ordered pattern, causing one or more Bragg peaks at low $q$ in equilibrium. In our case, the barrier is not strong enough to cause this microphase separation, as observed by the disordered structure of the voids and tunnels, and the height of the prepeak. The system is thus an amorphous fluid, with frustrated liquid-gas separation, but will possibly separate in microphases at higher attraction strength. This is similar
to other disordered phases showing incipient microphase separation, such as microemulsions \cite{microemulsions}.

In the following sections we make use of the density autocorrelation function, or normalized intermediate scattering function: 

\begin{equation}
\Phi_q(t)=\frac{1}{N} \left\langle \sum_{i=1}^N \sum_{j=1}^N \exp \left\{ i {\bf q}\cdot({\bf r}_i(t)-{\bf r}_j(0))\right\} \right\rangle/S(q) \label{phiq}
\end{equation}

\noindent where $N$ is the number of particles in the system, and ${\bf q}$ is the wavevector. The self part of it, $\Phi_q^s(t)$, is calculated by restricting the double summation to the case $i=j$. The brackets imply ensemble average, which we have calculated using different time origins as well as different samples, as explained above.

\subsection{MCT analysis}
\label{MCT}
Within mode coupling theory, the main region of the arrest line that represents an attraction-driven (gelation-like) transition corresponds to a so-called $A_2$ bifurcation \cite{dawson01}. The same is true for a standard, caging driven transition; but some differences with respect to the latter are predicted in the non-universal parameters of the bifurcation, such as the nonergodicity parameter. Such differences arise from the different driving mechanisms: caging by steric hindrance for the repulsive glass, and formation of a bonded network in attractive glass. At high density and attraction strength, where the repulsive and attractive glass lines meet, a high order singularity is found for short range potentials, characterized by a logarithmic decay in the autocorrelation functions \cite{dawson01,gotze02,puertas02,pham04}.

\begin{figure}
\psfig{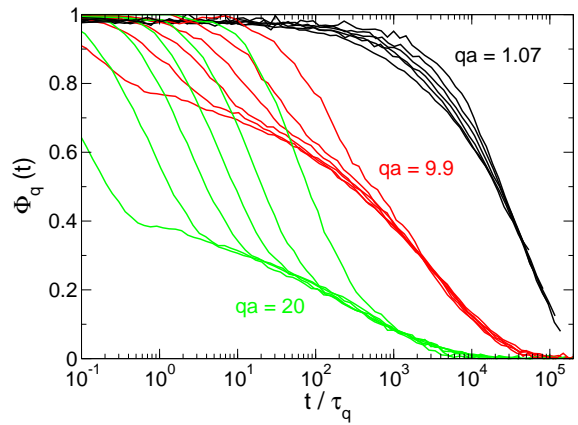}
\caption {\label{scaling} Time-rescaled density autocorrelation functions for $\phi_p=0.39$, $0.40$, $0.405$, $0.41$, $0.415$ and $0.42$.}
\end{figure}

Fig. \ref{scaling} shows the density autocorrelation functions for different states on the isochore $\phi_c=0.40$ approaching the attractive glass transition, for different wavevectors: at the pre-peak ($qa=1.07$) and two higher wavevectors. The correlators in Fig. \ref{scaling} have been time-rescaled to give collapse in the final $\alpha$-decay. Whereas such collapse is satisfactory for high wavevectors, at the pre-peak such a scaling is not possible: only at the final stages of decay can the curves can be said to overlap, as shown. For the master decay curves, observed at high $q$ where satisfactory scaling is accomplished, the decay of the density correlators is very stretched, and clear plateaus between the short-time $\beta$ relaxation and the terminal $\alpha$ region are thus not observed. Nevertheless, the values of the nonergodicity parameters $f_q$, estimated from the heights of the ill-defined plateaus that can be seen, are clearly much bigger than those of the glass transition in hard spheres \cite{voigtmann04}. This was correctly predicted by MCT \cite{dawson01}. These results are qualitatively similar to, but quantitatively different from, the incoherent autocorrelators studied previously \cite{puertas03}.

The decay from the plateau of the density autocorrelation function can be described by the von Schweidler expression,

\begin{equation}
\Phi_q(t)\:=\:f_q\,-\,h_q (t/\tau)^b \left[1 - k_q (t/\tau)^b\right] + O(t^{3b}) \label{vS}
\end{equation}

\noindent where $f_q$ is the nonergodicity parameter, $b$ is the von Schweidler exponent, and $h_q$ and $k_q$ are amplitudes. The scaling presented in Fig. \ref{scaling} shows that these parameters are almost the same for all states approaching the transition, consistent with their state independence as required by the $\alpha$-scaling of MCT, with a change only in the time scale $\tau$. (This applies for wavevectors that are not too low.) A similar expression can be used to describe the self part of the intermediate scattering function \cite{puertas03}, $\Phi_q^s(t)$, with a different set of parameters $f_q^s$, $h_q^s$ and $k_q^s$. 

\begin{figure}
\psfig{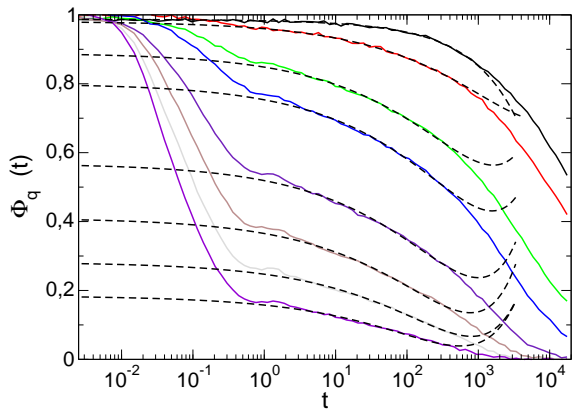}
\caption {\label{vonSchweidler} Density correlation functions for $\phi_c=0.40$ and $\phi_p=0.42$ for $qa=1.07$, $3.9$, $6.9$, $9.9$, $15$, $20$, $25$ and $30$. The fittings using eq. (\ref{vS}) are shown by the dashed lines. The von Scweidler exponent, $b$, is the same for all $q$, $b=0.37$.}
\end{figure}

Expression (\ref{vS}) is fitted to both the coherent and incoherent density autocorrelation functions; the fittings to the coherent correlation functions are presented in Fig. \ref{vonSchweidler} for various wavevectors. In all the fittings, the von Schweidler exponent was $b=0.37$, independent of the state or the wave vector. This figure shows that the decay of the density correlators indeed can be described using the von Schweidler decay.

\begin{figure}
\psfig{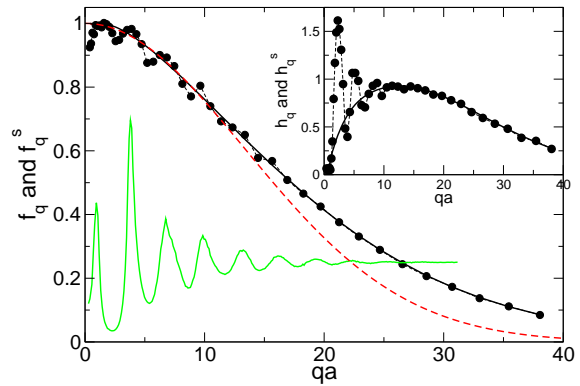}
\caption {\label{fq} Coherent and incoherent nonergodicty parameters (solid line and thick points, respectively). The structure factor for $\phi_p=0.42$ is also included and the Gaussian approximation fitting to $f_q^s$ (broken line). }
\end{figure}

The coherent and incoherent nonergodicity parameters, $f_q$ and $f_q^s$ respectively, and the amplitudes $h_q$ and $h_q^s$ are presented in Fig. \ref{fq}, with the structure factor for the state $\phi_p=0.42$. Whereas the incoherent nonergodicity parameter $f_q^s$ decays monotonically from $1$ to $0$ with increasing $q$, the coherent analogue $f_q$ oscillates visibly, in phase with $S_q$, for $qa \le 10$. At higher $qa$, both the coherent and incoherent nonergodicity parameters decay together, without oscillations. These results compare nicely with the theoretical predictions\cite{bergenholtz99}, and show that at high $q$ the collective and self dynamics are equivalent. But note that for short range attractions it is these high $q$ density fluctuations that dominate the arrest transition within MCT calculations. This is unlike the case of the repulsive glass, where the dominant fluctuations are around the main peak in $S(q)$, where the largest differences between the incoherent and coherent nonergodicity parameters arise. 

The amplitudes $h_q$ and $h_q^s$ are shown in the inset in Fig. \ref{fq}. The time scale $\tau$ in eq. (\ref{vS}) is needed in order to get absolute values of these amplitudes, but it cannot be accessed from the simulations. We have set $\tau$ to fulfil $\Phi_{q}(\tau_q)=f_q/e$ for $qa=9.9$. Then $h_q^s$ presents a maximum at $qa \approx 12$, whereas $h_q$ oscillates at low wavevectors and coincides with $h_q^s$ for large $q$. Contrary to $f_q$, the oscillations in the amplitude $h_q$ are out of phase with respect to $S(q)$, as predicted by MCT for other systems \cite{fuchs92} (forming repulsive glasses).

A Gaussian approximation to the van Hove function can be used \cite{fuchs99} to extract information from $f_q^s$. In this simple approximation, for low $q$, $\Phi_q^s \sim \exp\{-q^2 \langle \delta r^2 \rangle/6\}$, where $\langle \delta r^2 \rangle$ is the mean squared displacement. Thus, the incoherent nonergodicity parameter yields the localization length, $r_l$, via $f_q^s=\exp\{-q^2 r_l^2/6\}$ (dashed line in Fig. \ref{fq}). The localization length so obtained is $r_l=0.129a$ ($r_l^2=0.0168a^2$), about half the size of the interaction range, and much smaller than the localization length in the repulsion driven glass transition (which is of the order of the Lindemann distance, $r_l^2 \sim 0.1 a^2$). This finding shows that the driving mechanism of the arrest transition in our system is bond formation, i.e. density fluctuations at high $q$.

We now define $\tau_q$, the scaling time to collapse the $\alpha$-decay of the density correlator at a particular wavevector $q$, by the equation $\Phi_q(\tau_q)=f_q/e$. The results are plotted in Fig. \ref{tauq}. According to MCT, all the $\tau_q$ are governed by only one $q$-independent time scale of the $\alpha$-decay, $\tau$ in expression (\ref{vS}), which diverges at the transition point $\phi_p^G$ with a power law, while the self diffusion coefficient vanishes with a similar form:

\begin{figure}
\psfig{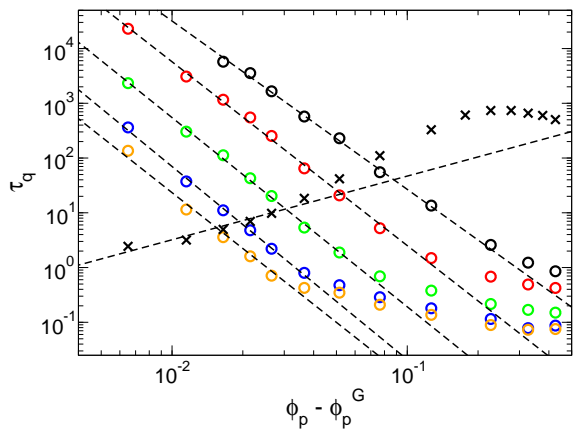}
\caption {\label{tauq} Time scale $\tau_q$ at $q=1.07$, $3.9$, $9.9$, $20$ and $30$ from top to bottom, and diffusion coefficient ($\times 10^3$), $D_0$, as a function of the distance to the transition. The dashed lines represent the power-law fittings according to (\ref{power-law}).}
\end{figure}

\begin{equation}
\tau \sim \left( \phi_p^G-\phi_p \right)^{-\gamma} \hspace{1cm} D_0 \sim \left( \phi_p^G-\phi_p \right)^{\gamma} \label{power-law}
\end{equation}

\noindent The exponent $\gamma$ is related to the von Schweidler exponent $b$ within MCT, leaving only one single parameter dependent on the interaction details. The transition point $\phi_p^G$ has been determined previously using the long-time scaling of the incoherent correlation functions \cite{puertas03}. Fig. \ref{tauq} shows that, with $\phi_p^G=0.4265$, $\tau_q$ indeed follows power-law divergences at all wavevectors, even for $qa=1.07$, where scaling was not found. The exponents for high wavevectors lie in the range $3.36 - 3.53$ without any trend, and $\gamma=3.08$ for $qa=1.07$. The time scales of the incoherent correlation functions also diverge following power laws, with exponents in the range \cite{puertas03} $3.1 - 3.25$.  Using the value of the von Schweidler exponent, $b=0.37$, the MCT relations yield $\gamma=3.44$, which lies well inside the range of observed exponents for the divergence of $\tau_q$. The similarity between all of the exponents found for the time scale shows that, as MCT predicts, the function $\tau_q(\phi_p)$ factorizes as $\tau(\phi_p) \omega(q)$, where $\omega(q)$ is a function only of $q$. Closer to the transition point, at $\phi_p=0.425$, the time scale deviates from the MCT power-law behaviour \cite{puertas03}, as also observed in other glass forming systems.

The diffusion coefficient, as obtained from the long time slope of the mean squared displacement, is also shown in Fig. \ref{tauq}. A power-law decrease is observed for $D_0$ with the same $\phi_p^G$, but with a smaller exponent, $\gamma=1.23$. Differences between these exponents ($\gamma$ determined from $\tau$ or $D_0$) have also been observed in simulations of other systems, such as Lennard-Jones particles\cite{kob95}, hard spheres\cite{doliwa00,voigtmann04}, polymer melts \cite{aichele01} or silica \cite{horbach01}, and imply the breakdown of the Stokes-Einstein relationship close to the glass transition. Alternatively, similar $\gamma$ exponents can be recovered for $D_0$ and $\tau$, so long as different transition point values $\phi_p^G$ are allowed in each fit; however this also represents a deviation from MCT predictions. Note that in any case, activated processes close to the transition will cause deviations from the MCT power-law predictions, restoring ergodicity beyond the transition point estimated by fitting to the power law region.

\begin{figure}
\psfig{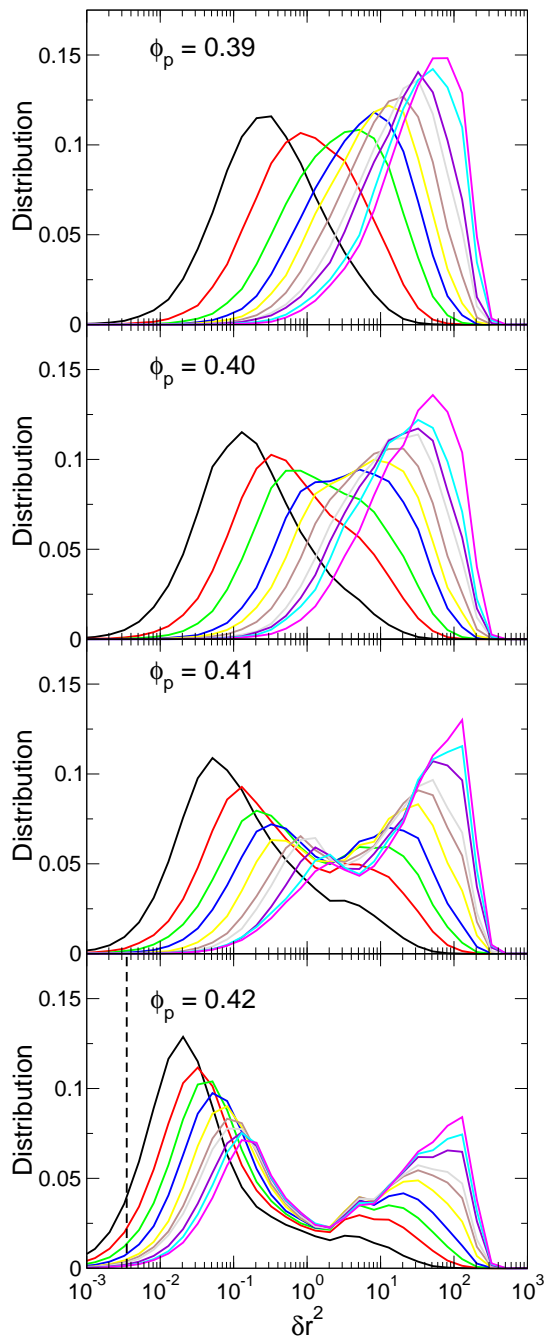}
\caption{\label{evolution} Distribution function $P(\delta r^2;t)$ of squared displacements for the states labeled, equispaced in time $t$. (These times are chosen such that $\langle \delta r^2 \rangle \approx 1 a^2$ for the shortest $t$ and $\langle \delta r^2 \rangle \approx 50 a^2$ for the largest; the actual $t$ values therefore vary with $\phi_p$.) As time evolves the distribution moves to the right. The thick blue lines correspond to $\langle \delta r^2 \rangle \approx 10 a^2$, where the distinction between fast and slow particles is made in the following figures. The dashed vertical line shows the squared displacement of a particle in the frozen environment.}
\end{figure}

The new results reported here, as well as previous analysis of the incoherent correlation functions \cite{puertas02,puertas03}, show that most (though not all) of the predictions drawn from the mode coupling theory are confirmed by our simulations. However, the presence of both structural and dynamical inhomogeneities would lead us to a different view of the system, less homogeneous and well behaved, which might not be reflected in averaged properties such as the coherent and incoherent correlators.

\subsection{Dynamical Heterogeneities}
\label{DH}
In Fig. \ref{evolution}, the probability distribution $P(\delta r^2;t)$ of squared displacements $\delta r^2$ measured over a time interval $t$ is presented (for various $t$ values), in a series of different states approaching the gel transition. (Curves with the same colors in different plots correspond to similar mean squared displacements.) In a homogeneous fluid, far from any arrest transition, this distribution is single peaked at all times; the peak moves to larger distances and broadens, with both mechanisms controlled by the diffusion coefficient \cite{puertas04}. This behavior is seen at the lowest polymer fraction studied here (upper panel in Fig. \ref{evolution}).

However, as the polymer fraction is increased, and the gel transition approached, the distribution becomes wider and, at $\phi_p=0.42$, two peaks become clearly visible, indicating two populations of particles with different mobilities (at least, on the time scales set by the intervals $t$). In fact, the peak corresponding to `slow' particles can already be noticed at $\phi_p=0.40$ as a shoulder, and at $\phi_p=0.41$ as a peak at lower displacements than the main peak, but  still moving to the right. The slow peak at $\phi_p=0.42$ is, however, at distances of the order of the localization length, showing a set of particles stuck at that distance. Note that the main peak, corresponding to a population of fast particles, is at squared distances three orders of magnitude larger than this by the final dataset. Even at short times, when $\langle \delta r^2 \rangle \approx 1 a^2$, a shoulder at large distances indicates the existence of a population of fast particles \cite{puertas04}. 

For high polymer fractions, a minimum develops at $\delta r^2 \sim a^2$. Once this happens, it is a meaningful approximation to describe the system with a two population model in which fast and slow particles are distinguished. At lower polymer fractions, however, this distinction is less clear, and only indicative.

\begin{figure}
\psfig{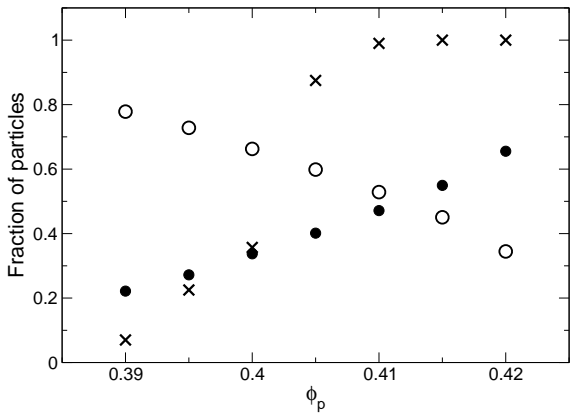}
\caption{\label{percolation} Number of fast (open circles) and slow (closed circles) particles, and percolation probability (crosses) as a function of the polymer fraction.}
\end{figure}

The evolution of the distributions presented in Fig. \ref{evolution} indicates that, as $\phi_p$ is raised, not only does the distinction between fast and slow particles become clearer, but also the fraction of slow particles increases \cite{puertas04}. This is studied in Fig. \ref{percolation}, where the fraction of slow and fast particles is presented as a  function of the polymer fraction for a fixed time, $t^*$, chosen such that $\langle \delta r^2(t^*) \rangle \approx 10 a^2$. (The distinction between fast and slow particles is made by slicing the distribution at $\delta r^2 = a^2$; see above.) Interestingly, at the highest $\phi_p$, the slow particles amount to 65\% 
of the system, and the square displacement at the peak in the distribution describing them increases only to $\sim 0.1 a^2$ on the time scales studied; this is in effect, a population of particles that remain stuck, very close to their initial spatial locations. 

However, the collective motion of the particles, even of the slow particles, can be observed by studying the mean squared displacements of one single particle moving in an environment of frozen particles. (In the simulations, only that particle is moved, using the configuration of the equilibrated system at the desired state \cite{puertas03}.) This displacement is constant up to very long times and is shown in the lowest panel of Fig. \ref{evolution} by a vertical dashed line. Its value is much smaller than the peaks of both the fast and the slow particles, indicating that although the slow particles are practically fixed, their cooperative motion allows them to explore longer distances than their cage size would be if the environment were frozen.

\begin{table}
\begin{ruledtabular}
\begin{tabular}{lccccc}
 $\phi_p$ & $n_{ave}$ & $n_f$ & $n_s$ & $n_{ff}$ & $n_{ss}$ \\ \hline
0.42 & 7.52 & 6.22 & 8.32 & 4.05 & 7.00 \\
0.415 & 7.31 & 6.61 & 8.19 & 4.50 & 6.36 \\
0.41 & 7.22 & 6.67 & 8.10 & 4.90 & 5.86 \\
0.405 & 7.15 & 6.74 & 8.07 & 5.17 & 5.43 \\
0.40 & 7.08 & 6.78 & 8.03 & 5.40 & 4.87 \\
0.395 & 6.99 & 6.75 & 7.98 & 5.56 & 4.56 \\
0.39 & 6.89 & 6.79 & 7.79 & 5.77 & 3.58 \\
\end{tabular}
\end{ruledtabular}
\caption{\label{table1} Mean number of neighbors for the whole system, $n_{ave}$, for the fast particles $n_f$ and slow particles $n_s$, fast neighbors of fast particles, $n_{ff}$ and slow neighbors of slow particles, $n_{ss}$. Note that the simulations studied here are longer than those presented previously\cite{puertas04}, making the results here more reliable.}
\end{table}

We now extend further the analysis of fast and slow distributions, aiming to understand better the nature of these two populations. In table \ref{table1} the mean number of neighbors of the fast and slow particles is presented, along with the average number of neighbors for the whole system. (`Neighbors' are defined as particles connected by a bond, i.e., lying within range of the attractive part of the potential.) The slow particles have more neighbors than average for all states, whereas the fast particles have less neighbors. Thus, we may conclude that the slow particles are in the inner parts of a bonding network in which both types of particle participate (the `skeleton' of the network), while the fast particles are in the outer layers of the network (its `skin'). As the gel transition is approached the difference between the two types of particles is more important, as reflected in their number of neighbors, thus implying that this picture of skeleton and skin is more appropriate closest to the gel. Note that the number of neighbors of the fast particles decreases as $\phi_p$ increases. Also in table \ref{table1} the number of fast neighbors of a fast particle, $n_{ff}$, and slow ones of a slow particle, $n_{ss}$, are given for all states. Supporting the simple picture of skeleton and skin, at high polymer fractions most of the neighbors of a slow particle are slow, and most of those of a fast particle are fast. Using the number of neighbors of every particle as a measure of the local density, this implies spatial correlations for both populations establishing further the link between structural heterogeneity and dynamic heterogeneity\cite{puertas04}.

Since the slow particles are mostly bonded to each other, and they can amount to more than half the system, it is possible that the slow particles percolate, forming a long lived elastic structure: truly a `skeleton' of the gel. The percolation probability for the slow particles (i.e. the probability of finding a cluster of slow particles, percolating in all three dimensions, for a given realisation of the system of the size that we use) is presented in Fig. \ref{percolation} as a function of the polymer fraction. This probability rises at $\phi_p \approx 0.40$, and percolation is systematically observed for $\phi_p \geq 0.41$. Interestingly, percolation of the slow particles sets in over just the same range of $\phi_p$ as marks the onset of a clear distinction between the slow and fast particles in the distribution of squared displacements (see Fig. \ref{evolution}). However, in the following we show that any link between these effects is somewhat indirect.

\begin{figure}
\psfig{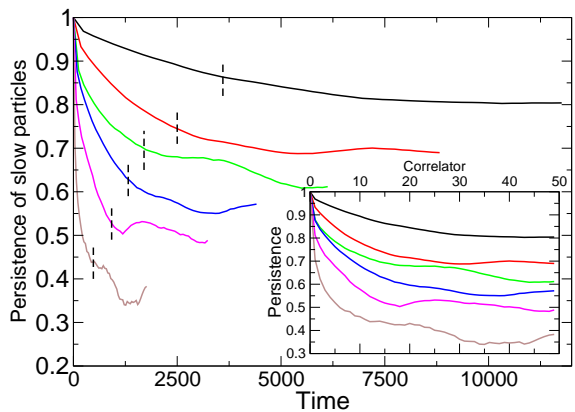}
\caption{\label{persistence} Fraction of slow particles that were slow in the first correlator as a function of time and the correlator index (inset). See text for the latter. From left to right: $\phi_p=0.39$, $0.40$, $0.405$, $0.41$, $0.415$ and $0.42$. The vertical dashed lines indicate $t^*$.}
\end{figure}

The picture of skeleton and skin presented above resembles the channel diffusion of sodium atoms in silica melts \cite{horbach02,jund01}. Channels in the Si-O network allow diffusion of the sodium atoms, resulting in a much higher mobility of the latter than for silicon or oxygen. The different dynamics is caused by the differences in the interactions, but this also results in structural heterogeneities, as in the case of gelation. However, in our case, there is no intrinsic difference between the two types of particles, fast and slow, and thus, in colloidal gelation, the stability of the two populations is not guaranteed, and an exchange between them may be ocurring. 

This exchange is addressed in Fig. \ref{persistence}, where the `persistence' of slow particles is presented. We define this persistence as follows \cite{puertas04}. We calculate a series of correlators with different initial times. A new correlator is initiated (alongside its predecessors) once the previous one has evolved to the point where $\langle \delta r^2 \rangle =a^2$; this ensures quasi-independence of one correlator from the next (because $a^2 \gg r_l^2$). The fast or slow character of every particle is decided after time interval $t^*$ for a given correlator, with $t^*$ chosen so that $\langle \delta r^2(t^*) \rangle \approx 10 a^2$. 
The persistence is defined as the fraction of particles, labelled as slow in a particular correlator, that remain slow when the classification is redone for a subsequent correlator. In the inset of Fig. \ref{persistence}, we show the persistence of the slow particles as a function of the correlator index; the main panel shows this as a function of elapsed time between correlators. (The inset effectively serves as a time-rescaling using the mean squared displacement.) As expected, the higher the polymer fraction, the slower the exchange between the two populations, and the slower the decay of the persistence of the slow particles. At low polymer fractions, most of the slow particles are reclassified as fast on a timescale of a few $t^*$ (a few correlators) and vice versa; but at high $\phi_p$, only a small fraction of slow particles become fast (presumably by escape from the percolating bonded network that they form) even at times that are an order of magnitude larger than the dynamical relaxation times $\tau_q$ set by the $\alpha$ relaxation time $\tau$ (see Fig. \ref{tauq}). These results are quite similar to those reported previously for the persistence of fast particles \cite{puertas04}.

The vertical dashed lines in Fig. \ref{persistence} mark the time when the mean squared displacement is equal to $10 a^2$, which is when the classification into fast and slow particles is made. This plot shows that the fraction of slow particles that become fast decreases with increasing $\phi_p$, although the absolute number of particles is between $90$ and $150$ for all the states. In the following we will analyse the dynamics of both sets of particles, fast and slow, on the basis of a simplified assumption that there is no exchange between them. Below $t^*$, this assumption is fairly accurate, but beyond $t^*$ the analysis becomes qualitative only.

\begin{figure}
\psfig{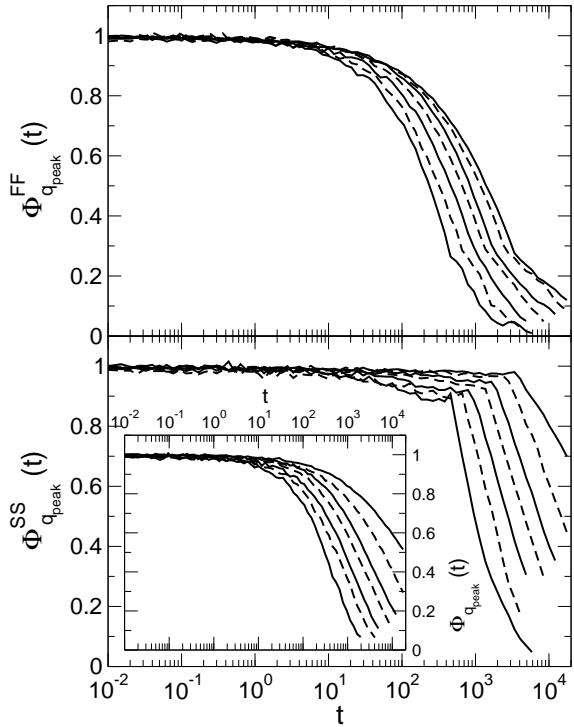}
\caption{\label{coherent} Intermediate scattering function for $\phi_p=0.39$, $0.395$, $0.40$, $0.405$, $0.41$, $0.415$ and $0.42$ from left to right. Contributions from fast-fast particles (upper panel), slow-slow particles (lower panel) and total function (inset).}
\end{figure}

In Fig. \ref{coherent}, we present the coherent intermediate scattering functions at the wavevector of the prepeak in $S(q)$, representative of structure at the length scale of any incipient microphase separation. The contributions from the fast particles (upper panel) and slow particles (lower panel), are shown, as well as the globally averaged functions (inset)\footnote{The contribution from a set of particles to the intermediate scattering function are calculated running the indices $i$ and $j$ in eq. \ref{phiq} to $N_{\alpha}$ and $N_{\beta}$, respectively, where $N_{\alpha}$ and $N_{\beta}$ are the number of fast, slow or all particles.}.
The time $t^*$ coincides closely with the kink in the curves, signalling the new mechanism entering the dynamics, i.e. exchange between the fast and slow populations. The contributions from the fast particles decay in all cases to below $0.2$ before $t^*$ is reached, while the contributions from the slow particles decay no lower than $0.85$ by the same time. Beyond $t^*$, the contribution from the slow particles decays much faster, possibly due to the conversion of some slow particles to fast. Similar behaviors were observed in the incoherent functions \cite{puertas04}, showing that the slow particles are tightly bonded to the network and their escape occurs only at very long times.

It should be noted that the contribution from the fast particles to both the coherent and incoherent functions decays slower for high polymer fractions, showing that these particles also feel the proximity of the transition (their bonds are longer lived). Thus, as the gel is approached, the slowing down of the system is due to all particles in the system, and not exclusively to the slow ones, whose fraction is increasing also. Also, it is interesting that no special feature is visible when the slow particles percolate, at $\phi_p=0.41$. This is evidence against any direct link between percolation of the slow particles and the onset of a clear classification (via the bimodality of $P(\delta r^2;t)$) into fast and slow populations.

These features are once again consistent with a skeletal network of stuck particles, surrounded by a skin of movable particles. The network is very stable in time, and relaxes on a time scale much larger than that of autocorrelators averaged over the whole system. However, the above results do not show whether this relaxation is a structural one, or caused by single particles leaving the network (thus converting slow particles into fast ones). Indeed, our definition of fast and slow particles is based only on the squared displacement of every particle, and thus it does not distinguish between structural relaxation and single particle motions. Therefore, taking the distribution of squared displacements at different times (different $t^*$) does provide qualitatively the same results, i.e. the correlation function from the slow particles hardly decays until $t^*$, when a rapid decay starts (only the fraction of fast and slow particles varies for different $t^*$).

\begin{figure}
\psfig{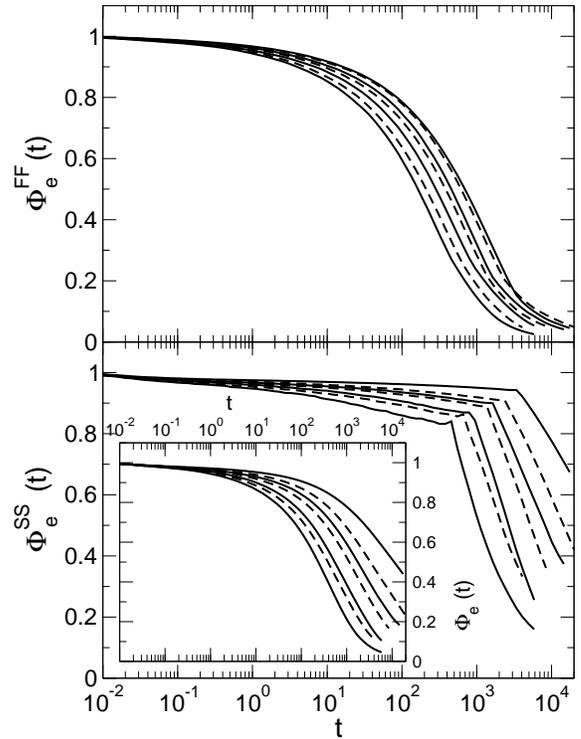}
\caption{\label{environment} Environment correlation function for the same states as Fig. \ref{coherent}. Contributions from the fast neighbors of fast particles (upper panel), slow neighbors of slow particles (lower panel= and total function (inset) are presented.}
\end{figure}

To clarify the origin of the long-time relaxation of the structure of the slow particles, we have calculated an {\sl environment} correlation function, $\Phi_e(t)$. By calculating the fraction of neighbors at given time, $t$, that were neighbors at time $t=0$, we quantify how the environment of the particles changes \cite{puertas04}. This function is similar to the cage correlation function, introduced by Rabani et al. \cite{rabani97} or the bond correlation function defined by Luzar and Chandler \cite{luzar93}, but the finite range of the attraction in our case provides an unambiguous way to define neighbors. In Fig. \ref{environment} we present $\Phi_e(t)$, and its contributions from the fast neighbors of fast particles, and the slow neighbors of the slow ones. These contributions are the dominant ones close to the gel, according to the distribution of neighbors presented in Table \ref{table1}. The data in Fig. \ref{environment} extend those reported previously \cite{puertas04} in both time and composition.

As expected the total environment correlation function decays more slowly at high polymer fractions, signalling the slower relaxation of the environment of the particles. As already observed in Fig. \ref{coherent}, the slow particles hardly change their environment before $t^*$, whereas the fast ones retain less than $20\%$ of their neighbors on this time scale; this is again consistent with our `skeleton and skin' description of the system. At larger times, however, the neighborhood of the slow particles finally relaxes (see Fig. 9) implying that the particles change their neighbors, thus escaping the network. This seems to suggest that the relaxation of the network of slow particles observed at long times occurs in a dissolution-like way, rather than by collective structural relaxation (which would allow slow regions to become fast {\sl without} reassignment of neighbors). Such a conclusion must be treated cautiously though, since our definition of neighbors (via bonding) is a strict one: a particle could alter its neighbors without structural reorganization on the cage scale or above.

\begin{figure}
\psfig{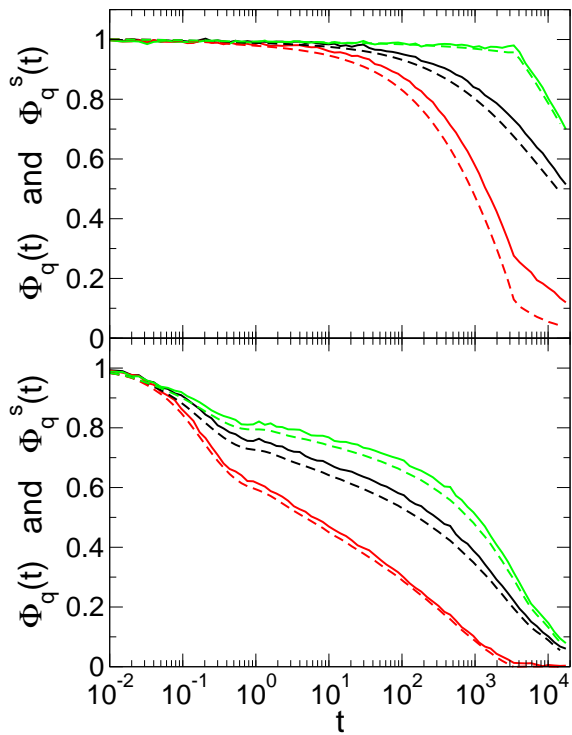}
\caption{\label{coherent-self} Coherent (continuous lines) and incoherent (broken lines) density correlation functions for $\phi_p=0.42$, and for $qa=1.07$ (upper panel) and $qa=9.9$ (lower panel). From top to bottom, contribution from the slow particles, whole system, and contribution from the fast particles.}
\end{figure}

Finally, we study the behaviour of the system at short distances by means of the density correlation function at high wavevector, $qa=9.9$. In Fig. \ref{coherent-self} both the coherent and incoherent intermediate scattering functions are presented for $qa=1.07$ (upper panel) and $qa=9.9$ (lower panel), at $\phi_p=0.42$. Again the contributions from the fast and slow particles to both correlation functions are shown. The correlation functions at this higher wavevector show that the fast particles are very mobile, but the slow ones are more trapped. However, as Fig. \ref{evolution} already showed, they are not completely stuck, and a significant decay of $\Phi_q$ is observed before $t^*$ for the slow particles (note that this contribution at low $q$ decays only to $0.95$). The bond correlation functions, not shown, indicate that $40\%$ of the bonds between slow particles break before $t^*$. Thus, the decorrelation at high $q$ implies bond breaking, and not only structural relaxation below $t^*$, although the environment of the particles does not change significantly (see Fig. \ref{environment}). 

\section{Conclusions}
\label{conc}
In this paper and its predecessors \cite{puertas02,puertas03,puertas04} we have presented the analysis of states close to gelation, in a system with short range interactions. Recent MCT results indicate that gelation at high density (as considered here) should be viewed as an attraction-driven glass transition; our system has been analysed to test these theoretical predictions. On the other hand, dynamical heterogeneities, although also observed in other glass forming systems, are found to be very pronounced in our system, and this might be taken as evidence against the MCT picture. We have studied these heterogeneities in some detail in an attempt to resolve their character.

The analysis of our system within MCT shows that most of the globally averaged properties, as measured by coherent and incoherent correlators, are correctly predicted. The $\alpha$-scaling of the density correlation function works for wavevectors that are not too low, and the system-universality of the von Schweidler exponent $b$ and the dynamic exponent $\gamma$, and the predicted relationship between these two, are confirmed. The driving mechanism, namely bond formation, causes differences between the attraction-driven glass studied here and and caging-driven repulsive glasses: the decay from the plateau is more stretched, the nonergodicity parameters are higher, and the localization length is shorter. All these trends are fully predicted by MCT. However, some significant differences from MCT predictions are noticed, such as the appearence of different exponents in fits for the divergence of $\tau_q$ and $1/D_0$, and the breakdown of the $\alpha$-scaling at low $q$. (The latter might be specifically related to the repulsive barrier used in our potential.)

By studying the distribution of squared displacements at fixed time in the system, it is possible to recognize two populations of particles with different mobility, in systems close to the gelation transition. The analysis of the correlation functions at different wavevectors (as well as environmental and bond correlators) shows that the slow particles form a quasi-frozen structure, which percolates at high enough $\phi_p$. (However the threshold is discernably lower than $\phi_p^G$ which, according to the MCT fits, marks the onset of gelation itself.) The fast particles are in the surface or `skin' of this cluster, and can escape more easily than the slow particles in the inner parts of it (the `skeleton'). There is a fairly clear correlation between the classification into fast and slow particles and the presence of structural heterogeneity as quantified in the local statistics of bonded neighbors. 

The slow particles, on the other hand, are not completely stuck, but can break their bonds and move over short distances. Despite this, they form a structure which hardly changes on lengthscales above that of a single particle. (In other words, broken bonds among slow particles are replaced by others before significant rearrangements can occur at larger length scales.) Only at very long times does the skeleton of slow bonded particles relax significantly its shape; and this seems to be due to a very slow exchange between fast and slow particles. This relaxation appears to be dissolution-like, in the sense that single slow particles escape to become fast, while elsewhere the reverse process occurs. However, evidence for cooperativity in the motion of fast and, more importantly, slow particles, is also found: the squared displacements of all particles is larger than that of single particles in the frozen environment.

Within our results, we cannot discern whether all of the slow particles will eventually change to fast, causing a complete relaxation of the structure as one would expect in a fluid phase. Although the curves in Fig. \ref{persistence} appear to saturate in all cases, the states at low $\phi_p$ must continue the exchange, since these are certainly fluid phases (see Fig. \ref{evolution}). Moreover, even our highest polymer density $\phi_p = 0.42$ is also a fluid phase if the gel transition is accurately identified \cite{puertas03} from the best von Schweidler fit ($\phi_p^G = 0.4265$), in which case the same can be expected for all our samples. However, the very presence of persistent structural and dynamical heterogeneity that endures through the entire lifetime of our simulation runs makes an accurate identification of the gel transition unreliable (to say nothing of issues of equilibration as raised in the introduction). This rather complicated physics arises despite the apparently `good behavior' of the globally averaged correlators -- which do seem to behave in line with MCT predictions for systems close to arrest (but still within the ergodic fluid phase).

The intriguing relation between the onset of percolation among the slow particles and the onset of a clear bimodality in $P(\delta r^2;t)$ among fast and slow populations of particles merits further study. No direct link can be established on the basis of our results, but it may be more than coincidence. The formation of a percolating skeletal cluster of slow particles at high enough $\phi_p$ could be instrumental in suppressing diffusive contributions that would otherwise arise from collective motions of finite clusters, thus sharpening considerably the bimodality of the population-level dynamics. However, one might expect this process also to imply a visible feature in the behavior of the correlation functions at low $qa$, which is not observed.

The somewhat complicated dynamics exhibited by our systems may caution against too sharp a dichotomy between different modelling strategies for glasses, at least when applied to colloidal suspensions \cite{paris}. If MCT is taken as {\sl a priori} incompatible with dynamic heterogeneity of any kind, then our results suggest that MCT's recent quantitative prediction of the experimental behavior of attracting colloids must in part be fortuitous. And in fact, qualitative retrodiction of some of these results has recently been achieved using facilitated dynamics models \cite{reichman}. But note also that very recent MCT work claims anyway to encompass DH \cite{MCTDH}; and a strong case has been made for a diverging dynamical lengthscale with associated fluctuations within MCT \cite{bouchaud}.

In any case the DH we observe is of discernibly different character to that seen in repulsion-driven glasses. The latter has formed part of the backdrop to recent developments in lattice-based models of DH involving facilitated dynamics \cite{chandler}. A fuller understanding of DH in dense colloids with short-range attractions must include some of these same elements, but may also involve an interplay between arrest and incipient phase separation (or, in our simulations, incipient microphase separation) which has long been argued to play a strong role in colloidal gelation at low density \cite{kroy03,poon}. 

\begin{center}
{\sc Acknowledgments} 

\end{center}
This work was funded in part under EPSRC Grant GR/S10377/01. A.M.P. acknowledges the financial support by the CICYT (project MAT2003-03051-CO3-01).

\end{document}